\documentclass[12pt]{article}
\usepackage{epsfig}
\usepackage{amssymb}
\usepackage{amsmath}
\usepackage{amsfonts}

\newcommand{\cH}{\mathcal{H}}

\newcommand{\be}{\begin{equation}}
\newcommand{\ee}{\end{equation}}
\newcommand{\bea}{\begin{eqnarray}}
\newcommand{\eea}{\end{eqnarray}}

\newcommand{\kt}{\rangle}
\newcommand{\br}{\langle}

\newcommand{\ed}{\end{document}}

\newcommand{\pbr}{\prec\!}
\newcommand{\pkt}{\!\succ}

\newcommand{\bi}{\begin{itemize}}
\newcommand{\ei}{\end{itemize}}

\begin{document}

\title{Comment on ``Reply to Comment on Time-dependent Quasi-Hermitian Hamiltonians
and the\\ Unitary Quantum Evolution''}
\author{\\
Ali Mostafazadeh
\\
\\
Department of Mathematics, Ko\c{c} University,\\
34450 Sariyer, Istanbul, Turkey\\ amostafazadeh@ku.edu.tr}
\date{ }
\maketitle

\begin{abstract}

I point out that if one defines the operator $U_R(t)$ as done by
M.~Znojil in his reply [arXiv:0711.0514v1] to my comment
[arXiv:0711.0137v1] and also accepts the validity of the defining
relation of $U_R(t)$ as given in his paper [arXiv:0710.5653v1],
one finds that the time-evolution of the associated quantum system
is not governed by the Schr\"odinger equation for the Hamiltonian
operator $H$ but an operator $H'$ which differs from $H$ if the
metric operator is time-dependent. In the latter case this
effective Hamiltonian $H'$ is not observable. This is consistent
with the conclusions of my paper [Phys.~Lett.~B \textbf{650}, 208
(2007), arXiv:0706.1872v2] which allow for unitary time-evolution
generated by unobservable Hamiltonians.

%\noindent Keywords:

\end{abstract}

%\tableofcontents
%\textheight = 22cm \topskip = -1cm \topmargin = -1cm

In \cite{r} M.~Znojil explains that my comment \cite{c} on his
paper \cite{znojil} is not relevant, because it relies on the
relations
        \be
        i\hbar\partial_t U_R(t)=H(t)U_R(t),~~~~U_R(0)=I,
        \label{U=}
        \ee
which he identifies as an ``\emph{incorrect}'' assumption. To
clarify the matter, recall that $U_R(t)$ was initially introduced
in Eq.~(14) of \cite{znojil} which reads
    \be
    |\Phi(t)\kt=U_R(t)|\Phi(0)\kt,
    \label{e1}
    \ee
where $|\Phi(t)\kt$ is an arbitrary evolving state vector.

Clearly (\ref{U=}) follows from (\ref{e1}), if one postulates the
standard Schr\"odinger time-evolution defined by the Hamiltonian
operator $H$, namely
        \be
        i\hbar\partial_t |\Phi(t)\kt=H(t)|\Phi(t)\kt.
        \label{sch-eq}
        \ee
This shows that in M.~Znojil's formulation the time-evolution is
not defined by the Hamiltonian $H$.

If one insists on defining $U_R(t)$ by the relation
    \be
    U_R(t)=\omega(t)^{-1}u(t)\omega(0),
    \label{e2}
    \ee
as done in \cite{r}, then (\ref{e1}) together with
$i\hbar\partial_t u(t)=h(t)u(t)$ and
$h(t)=\omega(t)H(t)\omega(t)^{-1}$ that are respectively listed as
Eqs.~(16) and (19) of \cite{znojil} yield
    \be
    i\hbar\partial_t |\Phi(t)\kt=\left[H(t)-i\hbar
    \omega(t)^{-1}\partial_t\omega(t)\right]|\Phi(t)\kt.
    \label{sch-eq-2}
    \ee
Therefore, in M.~Znojil's scheme, the evolving state vector is
determined not by $H(t)$ but by another operator namely
    \be
    H'(t):=H(t)-i\hbar
    \omega(t)^{-1}\partial_t\omega(t).
    \ee
It is easy to see that $H'$ coincides with $H$ if and only if the
metric operator is time-independent.

An important consequence of allowing a time-dependent metric
operator $\Theta$ in M.~Znojil's scheme is the
non-$\Theta$-pseudo-Hermiticity of $H'(t)$. This in turn means
that $H'(t)$ is not an observable. The situation becomes clear if
we recall the main result of \cite{paper} namely:
    \begin{itemize}
    \item[] \textbf{Theorem:} \emph{If the metric
operator is time-dependent, then the observability of the
Hamiltonian (the generator of the Schr\"odinger time-evolution) is
inconsistent with the unitarity of time evolution.}
    \end{itemize}
In M.~Znojil's scheme the generator of the Schr\"odinger
time-evolution, i.e., $H'(t)$, is not an observable. Therefore,
the unitarity of the Schr\"odinger time-evolution generated by
$H'(t)$ does not conflict with the above theorem.

Next, I would like to make two comments.
\begin{enumerate}
\item One can rewrite (\ref{sch-eq-2}) in the form
        \be
        i\hbar D_t |\Phi(t)\kt= H(t)|\Phi(t)\kt,~~~~~~
        D_t:=\partial_t+\omega(t)^{-1}\partial_t\omega(t),
        \label{sch-eq-3}
        \ee
    and identify $D_t$ with a covariant time-derivative. Such an
    approach has actually been pursued long ago in the context of
    quantum cosmology \cite{isham}.

\item A more straightforward formulation of the approach of
\cite{znojil} which does not involve any specific notation and is
mathematically unambiguous is as follows. Suppose that ${\cal H}$
is a reference Hilbert space with inner product
$\br\cdot|\cdot\kt$, $\Theta:{\cal H}\to{\cal H}$ is a possibly
time-dependent (positive) metric operator, $\cH_{\rm phys}$ is the
Hilbert space with the same vector space structure as ${\cal H}$
and the inner product
    \be
    \pbr\cdot,\cdot\pkt:=\br\cdot|\Theta\cdot\kt,
    \label{inn}
    \ee
and $H:{\cal H}\to{\cal H}$ be a possibly time-dependent
$\Theta$-pseudo-Hermitian operator. Suppose that the state vectors
$\Phi(t)$ evolve according to
    \be
    \Phi(t)=U(t)\Phi(0),
    \label{a1}
    \ee
where $U(t):{\cal H}\to{\cal H}$ is a linear densely-defined
invertible operator satisfying $U(0)=I$ and $I$ is the identity
operator. The unitarity of this evolution in the Hilbert space
$\cH_{\rm phys}$, i.e.,
    \be
    \pbr\Phi(t),\Phi(t)\pkt=\pbr\Phi(0),\Phi(0)\pkt,
    \label{uni}
    \ee
is equivalent to
    \be
    U(t)^\dagger\Theta(t)U(t)=\Theta(0),
    \label{a2}
    \ee
where $U(t)^\dagger$ is the unique operator satisfying
$\br\xi|U(t)^\dagger\zeta\kt=\br U(t)\xi|\zeta\kt$ for all
$\xi,\zeta\in{\cal H}$. So far $H(t)$ does not play any role in
this scheme. Let $H':{\cal H}\to{\cal H}$ be defined by
    \be
    H'(t):=i\hbar[\partial_t U(t)]U(t)^{-1}.
    \label{H-p}
    \ee
Then $\Phi(t)$ is a solution of the Schr\"odinger equation for a
Hamiltonian operator $H'(t)$,
    \be
    i\hbar\partial_t\Phi(t)=H'(t)\Phi(t).
    \label{sch-phi}
    \ee
Differentiating (\ref{a2}) and using (\ref{H-p}) in the resulting
equation, we also find
    \be
    {H'}^{\dagger}=\Theta(t)H'(t)\Theta(t)^{-1}+i\hbar
    [\partial_t \Theta(t)]\Theta(t)^{-1}.
    \label{H-p2}
    \ee
The Hamiltonian $H'$ is $\Theta$-pseudo-Hermitian, i.e., an
observable, if and only if $\Theta$ is constant. This argument is
independent of how one relates $H'(t)$ to $H(t)$, and it is in
complete agreement with the above theorem.

\end{enumerate}

In conclusion, M.~Znojil formulation of dynamics \cite{znojil}
allows for a unitary time-evolution with respect to a
time-dependent inner product, but this dynamics is generated by an
operator that is not an observable. If one uses the standard
notion of the ``Hamiltonian'' of a quantum system, namely as the
generator of the Sch\"odinger time-evolution, then the Hamiltonian
becomes unobservable. This is actually a direct implication of the
above theorem. What has been done in \cite{znojil} is to use the
term ``Hamiltonian'' for a different purpose. Such a non-standard
use of terminology is at the root of  M.~Znojil's disagreement
with the results of \cite{paper}.

\ed